\documentclass[a4paper,onecolumn,11pt,unpublished]{package}
\pdfoutput=1
\usepackage[utf8]{inputenc}
\usepackage[english]{babel}
\usepackage[T1]{fontenc}
\usepackage{amsmath}
\usepackage{hyperref}

\usepackage{tikz}
\usepackage{lipsum}

\usepackage{amsfonts}
\usepackage{amsthm}
\theoremstyle{plain}
\usepackage{algorithm}
\usepackage{algpseudocode}
\newtheorem{proposition}{Proposition}
\newtheorem{lemma}{Lemma}

\newtheorem{remark}{Remark}

\newtheorem{definition}{Definition}
\usepackage{bm}
\usepackage{braket}
\usepackage{float} 
\usepackage[numbers]{natbib}
\begin{document}

\title{Qudit Twisted-Torus Codes in the Bivariate Bicycle Framework}

\author{Mourad Halla}
%\affiliation{No institutional affiliation, Germany}
%\email{mourad.halla@desy.de}

\maketitle

\begin{abstract}
We study finite-length qudit quantum low-density parity-check (LDPC) codes from
translation-invariant CSS constructions on two-dimensional tori with twisted
boundary conditions. Recent qubit work [PRX Quantum 6, 020357 (2025)] showed
that, within the bivariate-bicycle viewpoint, twisting generalized toric
patterns can significantly improve the finite-size distance of qubit codes.
Building on this insight, we extend the construction and the search to qudit
codes over finite fields. Using algebraic methods based on Laurent-polynomial
ideals and Gr\"obner-basis computations over $\mathbb{F}_q$, we compute the
number of logical qudits on any twisted torus and identify compact codes with
favorable rate--distance tradeoffs. Overall, for the finite sizes explored,
twisted-torus qudit constructions typically achieve larger estimated
distances than their untwisted counterparts and, at the same number of
physical and logical qudits, larger estimated distances than previously
reported twisted qubit instances. The best new codes are tabulated.

\end{abstract}
%%%%%%%%%%%%%%%%%%%%%%%%%%
%%%%%%Introduction
%%%%%%%%%%%%%%%%%%%%%%%%%%
\section{Introduction}
\label{sec:intro}

Quantum error-correcting codes are essential for building large-scale
fault-tolerant quantum computers.  Among the many known families, topological
stabilizer codes play a special role by encoding quantum information into
global, nonlocal degrees of freedom while admitting local check operators.  The
paradigmatic example is Kitaev's toric code~\cite{KitaevToric}, which realizes
$\mathbb{Z}_2$ topological order on a square lattice and has inspired a broad
program of homological and topological quantum codes on more general lattices
and manifolds~\cite{BombinHomological}.

At the same time, quantum low-density parity-check (LDPC) codes have emerged as
a promising route toward scalable fault tolerance, since they retain
bounded-weight parity checks while allowing improved scaling of rate and
distance with system size~\cite{BreuckmannEberhardtLDPC}.  A foundational
algebraic construction is the hypergraph-product family of Tillich and
Z\'emor~\cite{TillichZemor}, and recent years have seen rapid progress on
high-performance LDPC families.  In particular, bivariate bicycle (BB) codes
combine bounded-weight checks with high decoding thresholds and substantially
higher encoding rates than surface-code families, while remaining
translation-invariant and not necessarily geometrically local~\cite{BravyiBB}.

A key technical ingredient behind several translation-invariant constructions is
an algebraic viewpoint, introduced by Haah~\cite{HaahModules}, which turns the
commutation relations and translation symmetry of stabilizer codes into
computations with finitely generated modules.  This perspective has enabled
systematic classification of topological order and efficient determination of
logical dimensions for broad classes of generalized Pauli stabilizer
codes~\cite{LiangTopOrder}.  It also clarifies how finite-size behavior can
depend strongly on the choice of periodic boundary conditions: for generalized
toric-type patterns, suitably \emph{twisted} torus identifications can change
the logical dimension at fixed blocklength and yield markedly better finite
codes than rectangular tori~\cite{LiangTwistedTori}.  Related ideas have been
adapted to planar geometries with open boundaries~\cite{LiangPlanar}.

Topological stabilizer codes can also be defined for qudits, i.e., local Hilbert
spaces of dimension $q>2$, and may offer practical advantages depending on the
hardware and noise model.  Examples include qudit surface codes~\cite{BullockBrennenQuditSurface}
and recent qudit LDPC constructions that generalize BB-type families~\cite{SpencerQuditLDPC}.
Motivated by these developments, we study generalized toric codes on twisted
tori in the qudit setting, extending the qubit twisted-torus construction of
Ref.~\cite{LiangTwistedTori} to qudits. We focus on prime local dimension $q$, so that
all algebra is carried out over the finite field $\mathbb{F}_q$. Using
Gr\"obner-basis computations in SageMath~\cite{SageMath}, we efficiently compute
the number of logical qudits for large collections of candidate codes, and we
estimate distances using the probabilistic algorithm implemented in the GAP
package \texttt{QDistRnd}~\cite{QDistRnd}. We perform a finite-length search
within a weight-$6$ ansatz and summarize the best instances found, together
with numerical distance estimates, in Tables~\ref{tab:3q4k}--\ref{tab:7q6k}.
%%%%%%%%%%%%%%%%%%%%%%%%%%%%%%%%%%%%%%%%%%%%%%%%%
%------------------------------------------------
%%%%%%%%%%%%%%%%%%%%%%%%%%%%%%%%%%%%%%%%%%%%%%%%%
%---------------------------------------------------------------------
\paragraph*{Relation to previous qubit work and contributions.}
Ref.~\cite{LiangTwistedTori} treats qubits, i.e.\ the case $q=2$, with all
algebra over $\mathbb{F}_2$.  The tools developed there, namely the
description of anyon sectors by the quotient ring
$\mathbb{F}_2[x^{\pm1},y^{\pm1}]/\langle f,g\rangle$, the resulting formula
for the number of logical qubits, and its finite-size version on twisted
tori, build on Refs.~\cite{HaahModules,LiangTopOrder}.  The present paper
extends these results to qudits of prime dimension $q$, with all algebra
over $\mathbb{F}_q$, and our contributions are the following.
(i)~For odd $q$ the $Z$-type generator of the qubit construction does not
commute with the $X$-type generator, because $-1\neq 1$ in $\mathbb{F}_q$; a
relative sign is needed (Lemma~\ref{lem:sign}).  This sign is invisible for
qubits and is the first point where the qudit case differs.
(ii)~For qudits each monomial of $f$ and $g$ carries a coefficient in
$\mathbb{F}_q^{\times}$.  This freedom does not exist for qubits; it changes
the ideal $\langle f,g\rangle$ and therefore both $k_{\max}$ and the
finite-size logical dimension (Remark~\ref{rem:coeff}).
(iii)~The anyon periods that decide which tori reach $k=k_{\max}$ are
multiplicative orders in extension fields $\mathbb{F}_{q^m}$, where $m$ is
the degree of the field extension, so for qudits they divide $q^{m}-1$
instead of the qubit value $2^{m}-1$
(Proposition~\ref{prop:period}).  As a result, the untwisted tori needed to
reach $k_{\max}$ can grow much faster with $q$, and twisted boundary
conditions become correspondingly more useful.
(iv)~We perform a search of this qudit family over $\mathbb{F}_3$,
$\mathbb{F}_5$ and $\mathbb{F}_7$ and report the best codes found
(Tables~\ref{tab:3q4k}--\ref{tab:7q6k}); the search procedure is described
in Appendix~\ref{app:algorithm}.

The rest of this manuscript is organized as follows. In
Sec.~\ref{sec:prelim_qudit} we review the qudit Pauli formalism and fix our
finite-field conventions. In Sec.~\ref{sec:algebraic_qudit_twisted} we
introduce the algebraic description of qudit CSS codes on twisted tori and
derive the formulas used to compute the logical dimension. In
Sec.~\ref{sec:numerics} we describe our numerical search and summarize the best
code instances found. Finally, in Sec.~\ref{sec:conclusion} we conclude with a
discussion of implications and future directions.
%%%%%%%%%%%%%%%%%%%%%%%%%%%%%%%%%%%%%%%%%%%%%%%%%
%------------------------------------------------
%%%%%%%%%%%%%%%%%%%%%%%%%%%%%%%%%%%%%%%%%%%%%%%%%
% [B] SECTION 2  (replaces the current Section 2)
%%%%%%%%%%%%%%%%%%%%%%%%%%%%%%%%%%%%%%%%%%%%%%%%%
\section{Qudit and finite-field preliminaries}
\label{sec:prelim_qudit}

In this work each physical degree of freedom is a qudit of Hilbert-space
dimension $q\ge 2$, and we use the symbol $q$ exclusively for this local dimension
throughout. We work with the standard generalized Pauli operators $X$ and $Z$
on a single qudit, defined on the computational basis
$\{\ket{j}\}_{j\in\mathbb{Z}_q}$ by
\begin{equation}
  X\ket{j}=\ket{j+1},\qquad Z\ket{j}=\omega^{\,j}\ket{j},\qquad
  \omega:=e^{2\pi i/q},
  \label{eq:qudit_XZ_def}
\end{equation}
so that
\begin{equation}
  ZX=\omega XZ .
  \label{eq:qudit_commutation}
\end{equation}
(The qubit case is recovered by setting $q=2$.)

Following Ref.~\cite{SpencerQuditLDPC}, we assume that $q$ is a prime power,
\begin{equation}
  q=p^s,\qquad p \text{ prime},\ s\in\mathbb{N},
  \label{eq:q_prime_power}
\end{equation}
and we carry out algebraic computations over the finite field $\mathbb{F}_q$.
When $s=1$ (i.e.\ $q=p$ is prime), $\mathbb{F}_q$ can be identified with
$\mathbb{Z}/q\mathbb{Z}$, so addition and multiplication are performed modulo
$q$:
\begin{equation}
  \mathbb{F}_q \cong \mathbb{Z}/q\mathbb{Z}.
  \label{eq:Fq_Zq}
\end{equation}
For $s\ge 2$ one may realize $\mathbb{F}_q$ as a degree-$s$ extension of
$\mathbb{F}_p$, for instance as $\mathbb{F}_p[u]/(\pi(u))$ with $\pi(u)$ an
irreducible polynomial of degree $s$. In this regime $\mathbb{Z}/q\mathbb{Z}$ is
not a field, so it cannot replace $\mathbb{F}_q$ in computations that require
multiplicative inverses.

In this paper we restrict to prime $q$.  For prime $q$, Haah showed that
every two-dimensional translation-invariant topological Pauli stabilizer
code is equivalent, by a constant-depth local Clifford circuit and up to
ancillas, to a finite number of copies of the $\mathbb{Z}_q$ toric code
\cite{HaahQudit}.  This classification supports the toric-code
interpretation underlying the anyon-counting arguments of
Refs.~\cite{LiangTopOrder,LiangTwistedTori}, whose algebraic formulation is
generalized here from $\mathbb{F}_2$ to $\mathbb{F}_q$; the points that are
specific to qudits are the sign structure of the generators, the
coefficient freedom, and the finite-field periodicities discussed in
Sec.~\ref{sec:algebraic_qudit_twisted}.  The prime-power case $s\ge2$
requires the $\mathbb{F}_q$-linear (Galois-qudit) stabilizer formalism
\cite{KetkarNonbinary} and is not studied here.

%%%%%%%%%%%%%%%%%%%%%%%%%%%%%%%%%%%%%%%%%%%%%%%%%
%------------------------------------------------
%%%%%%%%%%%%%%%%%%%%%%%%%%%%%%%%%%%%%%%%%%%%%%%%%
% [C] SECTION 3  (replaces the current Section 3)
%%%%%%%%%%%%%%%%%%%%%%%%%%%%%%%%%%%%%%%%%%%%%%%%%
\section{Qudit CSS codes on twisted tori: a Laurent-polynomial approach}
\label{sec:algebraic_qudit_twisted}

The Laurent-polynomial formalism for translation-invariant stabilizer codes was
introduced in Ref.~\cite{HaahModules} and further developed in
Refs.~\cite{LiangTopOrder, LiangTwistedTori, LiangPlanar}. In this work we apply it to
generalized toric codes on twisted tori in the qudit setting, extending the
corresponding qubit construction of Ref.~\cite{LiangTwistedTori}.

Throughout we consider qudits of prime dimension $q$ and work over the field
$\mathbb{F}_q$. We place two qudits in each unit cell and represent
translation-invariant Pauli patterns by their $X$- and $Z$-exponent data across
the unit cell. Lattice translations are encoded by commuting variables $x,y$,
so the natural coefficient ring is the Laurent polynomial ring
\begin{equation}
  R_q=\mathbb{F}_q[x^{\pm1},y^{\pm1}],
\end{equation}
and a translation by $(m,n)\in\mathbb{Z}^2$ acts by multiplication by the
monomial $x^m y^n$.

A translation-invariant CSS pattern is specified by two Laurent polynomials
$f(x,y),g(x,y)\in R_q$. We take bulk generators
\begin{equation}
A_v=\begin{bmatrix} f\\ g\\ 0\\ 0\end{bmatrix},
\qquad
B_p=\begin{bmatrix} 0\\ 0\\ \overline{g}\\ -\overline{f}\end{bmatrix},  
\qquad
\overline{x^m y^n}=x^{-m}y^{-n},
\label{eq:AvBp_qudit}
\end{equation}
where the bar denotes the antipode map, extended $\mathbb{F}_q$-linearly.

Commutation relations and syndromes are computed from the standard symplectic
form, with the antipode acting on the left factor:
\begin{equation}
v_1\cdot v_2:=\overline{v}_1^{\mathsf T}\Lambda v_2,\qquad
\Lambda=
\begin{bmatrix}
0&0&1&0\\
0&0&0&1\\
-1&0&0&0\\
0&-1&0&0
\end{bmatrix}.
\label{eq:sympl_qudit}
\end{equation}
For prime $q$, two Pauli operators commute if and only if the constant term of
$v_1\cdot v_2$ vanishes in $R_q$.

We can now explain the sign in \eqref{eq:AvBp_qudit}.  For qubits, i.e.\
over $\mathbb{F}_2$, one has $-1=1$, so the sign is invisible and
$B_p=(0,0,\overline g,\overline f)^{\mathsf T}$ is the usual choice
\cite{LiangTwistedTori}. For odd $q$ the sign is required, as the following
lemma shows.
\begin{lemma}[Commutation of the generators]
\label{lem:sign}
Let $q$ be an odd prime and let $f,g\in R_q$ be nonzero.  With the
symplectic form \eqref{eq:sympl_qudit}, all lattice translates of $A_v$ and
$B_p$ in \eqref{eq:AvBp_qudit} commute.  If the last entry of $B_p$ is
replaced by $+\overline{f}$, then
$A_v\cdot B_p=2\,\overline{f}\,\overline{g}\neq 0$, and some translate of
$A_v$ fails to commute with some translate of $B_p$.
\end{lemma}

\begin{proof}
$A_v$ is purely $X$-type and $B_p$ purely $Z$-type, so only the upper-right
block of $\Lambda$ contributes to \eqref{eq:sympl_qudit}:
$A_v\cdot B_p=\overline{f}\,\overline{g}+\overline{g}\,(-\overline{f})=0$,
since $R_q$ is commutative.  The translate of $B_p$ by $(m,n)$ is
$x^{m}y^{n}B_p$, and $A_v\cdot(x^{m}y^{n}B_p)=x^{m}y^{n}\,(A_v\cdot B_p)$,
whose constant term is the coefficient of $x^{-m}y^{-n}$ in $A_v\cdot B_p$
\cite{HaahModules}; hence the polynomial $A_v\cdot B_p$ vanishes if and only
if all translates commute.  With $+\overline f$ the same computation gives
$2\,\overline f\,\overline g$, which is nonzero because $2\neq0$ for odd $q$
and $R_q$ has no zero divisors, so $\overline f\,\overline g\neq0$ for
nonzero $f,g$.
\end{proof}

\noindent
In block form, \eqref{eq:AvBp_qudit} reads $H_X=[\,A\;|\;B\,]$ and
$H_Z=[\,B^{\mathsf T}\;|\;-A^{\mathsf T}\,]$, where $A$ and $B$ are the
matrices of multiplication by $f$ and $g$ on the torus.  This is the qudit
counterpart of the block form $H_Z=[\,B^{\mathsf T}\;|\;A^{\mathsf T}\,]$ of
qubit bivariate bicycle codes \cite{BravyiBB}; one has
$H_XH_Z^{\mathsf T}=AB-BA=0$ because $A$ and $B$ commute.

In the generalized toric family studied later we take $f$ and $g$ to be sparse:
\begin{align}
f(x,y) &= \sum_{i=1}^{t_f} l_i\, x^{a_i}y^{b_i}, \label{eq:ansatz_general_f}\\
g(x,y) &= \sum_{j=1}^{t_g} m_j\, x^{c_j}y^{d_j}, \label{eq:ansatz_general_g}
\end{align}
where $a_i,b_i,c_j,d_j\in\mathbb{Z}$ specify lattice displacements, and the
coefficients satisfy $l_i,m_j\in\mathbb{F}_q^\times$ (i.e.\
$l_i,m_j\neq 0$). We assume all monomials $x^{a_i}y^{b_i}$ (resp.\
$x^{c_j}y^{d_j}$) are distinct. The (bulk) check weight is then
\[
w=\mathrm{wt}(f)+\mathrm{wt}(g)=t_f+t_g,
\]
where $\mathrm{wt}(\cdot)$ denotes the number of nonzero monomials.

By an overall lattice translation and rescaling by a unit in $\mathbb{F}_q^\times$,
we may assume $(a_1,b_1)=(0,0)$ and $l_1=m_1=1$, so that
\begin{equation}
f(x,y)=1+\sum_{i=2}^{t_f} l_i x^{a_i}y^{b_i},\qquad
g(x,y)=1+\sum_{j=2}^{t_g} m_j x^{c_j}y^{d_j}.
\label{ansatz}
\end{equation}

\begin{definition}[Stabilizer range]
\label{def:range}
For $f,g$ as in \eqref{ansatz}, let $r_x$ and $r_y$ be the differences
between the largest and the smallest exponent of $x$ and of $y$,
respectively, over all monomials of $f$ and $g$.  The stabilizer range is
$r=r_x+r_y$.  It bounds the extent of the support of a check operator on
the lattice, and it is the quantity we minimize among codes with equal
parameters.  For example, $f=1+x^{-2}+x^{2}y^{-3}$ and
$g=1+y^{-3}+x^{3}y^{2}$ have $r_x=3-(-2)=5$, $r_y=2-(-3)=5$ and $r=10$; the
toric code $f=1+x$, $g=1+y$ has $r=2$.
\end{definition}

\begin{remark}[Coefficient freedom for $q>2$]
\label{rem:coeff}
For $q=2$ the only nonzero element of $\mathbb{F}_2$ is $1$, so every
coefficient equals $1$ and a pattern of given weight is fixed by its
exponents alone, as in the binary construction of
Ref.~\cite{LiangTwistedTori}.  For $q>2$ each non-constant monomial can carry
any coefficient in $\mathbb{F}_q^{\times}$.  With the normalization above
this gives $(q-1)^{\,t_f+t_g-2}$ coefficient patterns for each choice of
exponents, $(q-1)^{4}$ for the weight-6 family of Sec.~\ref{sec:numerics}.
The coefficients are not a cosmetic change: they enter the ideal
$\langle f,g\rangle$, and therefore the topological-order condition,
$k_{\max}$ and the finite-size logical dimension all depend on them.  A
coefficient different from $1$ appears in the check operator as a power
$X^{l}$ or $Z^{m}$ of a generalized Pauli operator.  This larger design
space is the second qudit-specific ingredient of our construction, and the
search of Sec.~\ref{sec:numerics} makes use of it.
\end{remark}

Using the symplectic form \eqref{eq:sympl_qudit} and the unit-cell basis vectors
$\mathbf{X}_1,\mathbf{X}_2,\mathbf{Z}_1,\mathbf{Z}_2$, the excitation map
$\epsilon$ (Pauli $\mapsto$ $(A_v,B_p)$-syndrome) is
\begin{equation}
\epsilon(\mathbf{X}_1)=[0,-g],\quad                                          
\epsilon(\mathbf{X}_2)=[0,f],\quad
\epsilon(\mathbf{Z}_1)=[\overline{f},0],\quad
\epsilon(\mathbf{Z}_2)=[\overline{g},0].
\label{eq:epsilon_qudit}
\end{equation}
(Overall signs of the syndromes play no role below, since they do not change 
the ideals generated.)

Let $\mathcal{S}$ denote the stabilizer module generated by all lattice
translates of $A_v$ and $B_p$.  We impose the standard topological-order
condition for translation-invariant CSS patterns by requiring the minimal
intersection of the ideals generated by $f$ and $g$:
\begin{equation}
\langle f\rangle \cap \langle g\rangle = \langle fg\rangle .
\label{eq:TO_ideal_qudit}
\end{equation}
Throughout we take $q$ to be prime and work over
$R_q=\mathbb{F}_q[x^{\pm1},y^{\pm1}]$.  When \eqref{eq:TO_ideal_qudit} holds,
the maximal logical dimension achievable on sufficiently large periodic
geometries is determined by the quotient $R_q/\langle f,g\rangle$,
\begin{equation}
k_{\max}
=2\,\dim_{\mathbb{F}_q}\!\left(\frac{R_q}{\langle f,g\rangle}\right).
\label{eq:kmax_qudit}
\end{equation}

Finite twisted tori are specified by three integers $(\alpha,\beta,\gamma)$ with
$\alpha,\beta>0$: we identify lattice sites under translations by the vectors
$(0,\alpha)$ and $(\beta,\gamma)$.  Here $\alpha$ and $\beta$ set the vertical
and horizontal periods, while $\gamma\in\mathbb{Z}$ is the \emph{twist}
parameter describing the vertical shift accumulated when wrapping once around
the second cycle.  The fundamental domain then contains $\alpha\beta$ unit
cells (hence $n=2\alpha\beta$ physical qudits in our two-qudits-per-cell
convention).  In the Laurent-polynomial language these identifications impose
the boundary relations
\begin{equation}
y^\alpha-1=0,\qquad x^\beta y^\gamma-1=0
\quad\text{in }R_q.
\label{eq:torus_relations}                                                    
\end{equation}
For fixed $(f,g)$, the corresponding twisted-torus code is described by the
finite-dimensional quotient
\begin{equation}
\mathcal{M}(\alpha,\beta,\gamma):=
\frac{R_q}{\langle f,g,\,y^\alpha-1,\,x^\beta y^\gamma-1\rangle},
\end{equation}
and the number of logical qudits is
\begin{equation}
k(\alpha,\beta,\gamma)
=2\,\dim_{\mathbb{F}_q}\mathcal{M}(\alpha,\beta,\gamma)
\le k_{\max}.
\label{eq:k_twisted_qudit_short}
\end{equation}

The field $\mathbb{F}_q$ enters through the periods of the anyons, i.e.\ of
the elements of $R_q/\langle f,g\rangle$ as in Ref.~\cite{LiangTwistedTori}.
These periods decide on which tori the bound $k\le k_{\max}$ is reached.
Since $R_q/\langle f,g\rangle$ is finite dimensional, the powers
$1,y,y^{2},\dots$ are linearly dependent modulo $\langle f,g\rangle$, so
$\langle f,g\rangle\cap\mathbb{F}_q[y]$ is a nonzero ideal of
$\mathbb{F}_q[y]$.  Let $h(y)$ be its monic generator; it is the univariate
element of a lexicographic Gr\"obner basis.  Define $h'(x)$ in the same way.
As $y$ is a unit in $R_q$, we may assume $h(0)\neq0$.

\begin{proposition}[Anyon periods over $\mathbb{F}_q$]
\label{prop:period}
Let $q$ be prime and let \eqref{eq:TO_ideal_qudit} hold.
\begin{itemize}
\item[(a)] $y^{L}-1\in\langle f,g\rangle$ if and only if $h(y)$ divides
$y^{L}-1$; the same holds for $x$ and $h'(x)$.
\item[(b)] If $h=\prod_i h_i$ with distinct irreducible factors $h_i$ of
degrees $m_i$, the smallest such $L$ is
\begin{equation}
L_y=\operatorname{lcm}_i\bigl(\operatorname{ord}h_i\bigr),
\label{eq:period_formula}
\end{equation}
where $\operatorname{lcm}_i$ denotes the least common multiple over all
factors and $\operatorname{ord}h_i$ is the order of a root of $h_i$ in
$\mathbb{F}_{q^{m_i}}^{\times}$, i.e.\ the smallest $t\ge1$ with
$\eta^{t}=1$ for a root $\eta$ of $h_i$.  Each $\operatorname{ord}h_i$
divides $q^{m_i}-1$.  In particular, if $h$ is irreducible of degree $m$,
then $L_y$ divides $q^{m}-1$.
\item[(c)] $k(\alpha,\beta,\gamma)=k_{\max}$ if and only if $y^{\alpha}-1$
and $x^{\beta}y^{\gamma}-1$ lie in $\langle f,g\rangle$.  For an untwisted
torus this means that $\alpha$ must be a multiple of $L_y$ and $\beta$ a
multiple of $L_x$, so the smallest untwisted torus with $k=k_{\max}$ has
$n=2L_xL_y$ qudits.
\end{itemize}
\end{proposition}

\begin{proof}
(a) $y^{L}-1$ involves only $y$, so it lies in $\langle f,g\rangle$ exactly
when it lies in $\langle f,g\rangle\cap\mathbb{F}_q[y]=\langle h\rangle$,
i.e.\ when $h$ divides it.
(b) $h$ divides $y^{L}-1$ exactly when every root of $h$ is an $L$-th root
of unity, i.e.\ when the order of every root divides $L$; the smallest such
$L$ is the least common multiple of these orders.  The roots of $h_i$ lie
in $\mathbb{F}_{q^{m_i}}$, whose multiplicative group has $q^{m_i}-1$
elements, so each order divides $q^{m_i}-1$ \cite{LidlNiederreiter}.
(c) If both relations lie in $\langle f,g\rangle$, adding them does not
change the ideal, so $k=k_{\max}$.  Conversely, if $k=k_{\max}$, the
quotient by $\langle f,g\rangle$ and the quotient by the larger ideal
$\langle f,g,y^{\alpha}-1,x^{\beta}y^{\gamma}-1\rangle$ have the same
dimension, so the two ideals coincide.  The untwisted case follows from
(a).
\end{proof}

\begin{remark}
If $h=\prod_i h_i^{e_i}$ has repeated factors, then
\[
L_y=\operatorname{lcm}_i\bigl(\operatorname{ord}h_i\bigr)\,
q^{\lceil\log_q\max_i e_i\rceil}.
\]
Indeed, writing $L=q^{r}L'$ with $L'$ not divisible by $q$, in
characteristic $q$ one has $y^{L}-1=(y^{L'}-1)^{q^{r}}$, while $y^{L'}-1$
has no repeated roots.  Hence a factor $h_i^{e_i}$ requires $q^{r}\ge e_i$
for all $i$, i.e.\ $r\ge\log_q\max_i e_i$, and the smallest admissible
$r$ is $\lceil\log_q\max_i e_i\rceil$, which gives the stated factor.  The
same applies to $h'(x)$.
\end{remark}

\noindent
Part (c) is Corollary~2 of Ref.~\cite{LiangTwistedTori}, proved there for
qubits by an argument that does not depend on the field.  Parts (a) and (b)
are the qudit version of Corollary~1 of Ref.~\cite{LiangTwistedTori}: the
base $2$ in the qubit bound $L\mid 2^{m}-1$ becomes $q$.  

%%%%%%%%%%%%%%%%%%%%%%%%%%%%%%%%%%%
%Numer
%%%%%%%%%%%%%%%%%%%%%%%%%%%%%%%%%%%
%---------------------------------------------
\section{Numerical results}
\label{sec:numerics}

In this section we numerically search for good finite-length qudit LDPC codes
obtained by placing translation-invariant generalized toric patterns on twisted
tori.  For each choice of bulk polynomials $(f,g)$ and twisted boundary data
$(\alpha,\beta,\gamma)$, we evaluate the number of logical qudits
$k(\alpha,\beta,\gamma)$ via the quotient in \eqref{eq:k_twisted_qudit_short}
(using Gr\"obner-basis computations over $\mathbb{F}_q$) and then estimate the
distance $d$ from the minimum weight of a nontrivial logical operator.  Our goal
is to realize codes with large $k$ and $d$ at modest blocklength
$n=2\alpha\beta$, and in particular to find twisted geometries that achieve
$k(\alpha,\beta,\gamma)$ close to the infinite-plane bound $k_{\max}$.

To keep the checks low-weight and the search space manageable, we restrict to
\emph{weight-6} bulk generators. Concretely, we take $\mathrm{wt}(f)=\mathrm{wt}(g)=3$,
so each bulk stabilizer has weight
$w=\mathrm{wt}(f)+\mathrm{wt}(g)=6$, and we use the sparse ansatz \eqref{ansatz}
\begin{equation}
f(x,y)=1+l_1 x^{a_1}y^{b_1}+l_2 x^{a_2}y^{b_2},\qquad
g(x,y)=1+m_1 x^{c_1}y^{d_1}+m_2 x^{c_2}y^{d_2},
\label{eq:weight6_ansatz}
\end{equation}
where $l_1,l_2,m_1,m_2\in\mathbb{F}_q^\times$ and
$a_i,b_i,c_i,d_i\in\mathbb{Z}$.

We summarize the best codes found in Tables~\ref{tab:3q4k}--\ref{tab:7q6k}, reporting $[[n,k,d]]_q$, the
bulk polynomials $(f,g)$, the twist vectors $\vec a_1=(0,\alpha)$ and
$\vec a_2=(\beta,\gamma)$, and the figure of merit $k d^2/n$, motivated by the
two-dimensional Bravyi--Poulin--Terhal tradeoff~\cite{TerhalBound1,TerhalBound2}.
The search over $(f,g)$ and $(\alpha,\beta,\gamma)$ uses Gr\"obner-basis
computations in SageMath~\cite{SageMath}. The code distance $d$ is estimated using a probabilistic algorithm implemented
in the GAP package \texttt{QDistRnd}~\cite{QDistRnd}; the resulting value
provides an upper bound on $d$.  During the search, distances were screened
with $3000$ information sets; for each $(n,k,q)$ the $20$ candidates with
the largest $kd^{2}/n$ were then recomputed with $30\,000$ information sets,
and the best of these is reported.

Overall, the best qudit twisted-torus instances attain larger values of
$k d^{2}/n$ than the corresponding twisted-torus qubit benchmarks, consistent
with larger distances $d$ at matched $(n,k)$ for the same check weight
$w=6$~\cite{LiangTwistedTori}. For example, at $(n,k)=(48,4)$ we obtain
$[[48,4,9]]_{q=3}$ with $k d^{2}/n=6.75$ compared to the qubit instance
$[[48,4,8]]_{q=2}$ with $k d^{2}/n=5.33$, and at $(n,k)=(114,4)$ we obtain
$[[114,4,16]]_{q=3}$ with $k d^{2}/n=8.98$ compared to $[[114,4,14]]_{q=2}$
with $k d^{2}/n=6.88$. The same trend persists at larger blocklengths: at
$(n,k)=(156,4)$ we find $[[156,4,19]]_{q=3}$ with $k d^{2}/n=9.26$
compared to $[[156,4,16]]_{q=2}$ with $k d^{2}/n=6.56$, and at
$(n,k)=(224,6)$ we find $[[224,6,23]]_{q=3}$ with $k d^{2}/n=14.17$
compared to $[[224,6,20]]_{q=2}$ with $k d^{2}/n=10.71$. Since $n$ and $k$
are fixed in these comparisons, the increase in $k d^{2}/n$ directly
reflects the larger distance; as described in Appendix~\ref{app:algorithm},
our distances are upper bounds obtained with QDistRnd.

Untwisted tori are the special case $\gamma=0$ of our search, so the best
code found over all boundary conditions is never worse than the best
untwisted code found in the same search. Moreover, for every code listed in
the tables we also evaluated the same pair $(f,g)$ on the untwisted torus
with the same $(\alpha,\beta)$. In all tabulated cases the selected code has
a nonzero twist, and whenever the untwisted instance also realizes the
target $k$, its distance is smaller. For these codes the improvement is
thus due to the boundary twist and not to a different choice of bulk
polynomials. For each $(n,k,q)$ the tables list the code with the largest
$kd^{2}/n$ found and, among ties, the one with the smallest stabilizer
range (Definition~\ref{def:range}).

%%%%%%%%%%%%%%%%%%%%%%%%%%%%%%%%%%%%%%%%
%%%%%%%%%table q=3 , k =4
%%%%%%%%%%%%%%%%%%%%%%%%%%%%%%%%%%%%%%%%
\begin{table}[htbp]
\centering
\caption{Best weight-$6$ twisted-torus qudit generalized toric codes found in our search with $q=3$ and $k=4$, listing $(f,g)$, twist parameters, and the performance metric $k d^{2}/n$.}
\label{tab:3q4k}
\begin{tabular}{c c c c c c}
\hline\hline
$[[n,k,d]]_{q=3}$ &
\begin{tabular}{@{}c@{}} $f(x,y) =$ \\  \end{tabular} &
\begin{tabular}{@{}c@{}} $g(x,y) =$ \\ \end{tabular} &
$\vec a_1$ &
$\vec a_2$ &
%$\dfrac{k d^{2}}{n}$ \\
$k d^{2}/n$\\
\hline
% -------------------
$[[42,4,8]]$ & $1 + y^{-2} + x^{-1} y^{-2}$     & $x^{2} y + 1 + x^{-1}y^{-3}$     & $(0,3)$ & $(7,-2)$ & $6.10$ \\
%%%%%%
$[[48,4,9]]$ & $-x^2 y + 1 - x^{-3} y^{-1}$ & $y + 1 + x^{-1} y$ & $(0,8)$ & $(3,-1)$ & $6.75$ \\
%%%
$[[64,4,11]]$ & $1 - x^{-1} y^{2} - x^{-3} y^{-2}$ & $y^{3} + 1 + x^{-1} y^{-3}$ & $(0,16)$ & $(2,-5)$ & $7.56$ \\
%%%%%
$[[78,4,12]]$ & $x^{2}y+xy^{2}+1$ & $x^{2}y^{3}+1+x^{-2}$ & $(0,13)$ & $(3,-3)$ & $7.38$ \\
%%%
$[[102,4,15]]$ & $x^{2} y + 1 + x^{-2} y^{-2}$ & $1 + y^{-2} + x^{-3} y^{-3}$ & $(0,17)$ & $(3,-8)$ & $8.82$ \\
%%%%%%%%
$[[114,4,16]]$ & $x^{2} y^{-3} + 1 + x^{-2}$ & $x^{3} y^{2} + 1 + y^{-3}$ & $(0,19)$ & $(3,-9)$ & $8.98$ \\
%%%%%
$[[138,4,18]]$ & $x^{3} y^{-1} + 1 + y^{-2}$ & $x^{2} y^{-1} + x y^{3} + 1$ & $(0,23)$ & $(3,-11)$ & $9.39$ \\
%%%%
$[[156,4,19]]$ & $y^{2}+1-x^{-3}y^{2}$ & $1-x^{-1}y^{3}+x^{-2}y^{3}$ & $(0,13)$ & $(6,-10)$ & $9.26$ \\
%%%%%
$[[174,4,21]]$ & $xy^{2}+1+x^{-1}y^{-3}$ & $x^{3}y^{-1}+1+x^{-3}y^{-2}$ & $(0,29)$ & $(3,-16)$ & $10.14$ \\
%%%%
$[[222,4,25]]$ & $x^{3}y^{3}+x^{3}y^{-3}+1$ & $x^{3}y+1+x^{-2}y^{-2}$ & $(0,111)$ & $(1,-13)$ & $11.26$ \\
%%%
$[[228,4,25]]$ & $-x^{3}y^{3}+y^{2}+1$ & $x^{2}y^{3}-xy^{-1}+1$ & $(0,19)$ & $(6,-8)$ & $10.96$ \\
%%%%
$[[240,4,26]]$ & $x^{2}y^{-2}+xy^{-3}+1$ & $-x^{2}+1+x^{-1}y^{-2}$ & $(0,24)$ & $(5,10)$ & $11.27$ \\
% ---------------------------------------------------------------
%\hline\hline
\end{tabular}
\end{table}
%%%%%%%%%%%%%%%%%%%%%%%%%%%%%%%%%%%%%%%%
%%%%%%%%%q=3, k=6
%%%%%%%%%%%%%%%%%%%%%%%%%%%%%%%%%%%%%%%%
\begin{table}[htbp]
\centering
\caption{Same as Table~\ref{tab:3q4k}, but with $k=6$.}
\label{tab:3q6k}
\begin{tabular}{c c c c c c}
\hline\hline
$[[n,k,d]]_{q=3}$ &
\begin{tabular}{@{}c@{}} $f(x,y) =$ \\  \end{tabular} &
\begin{tabular}{@{}c@{}} $g(x,y) =$ \\ \end{tabular} &
$\vec a_1$ &
$\vec a_2$ &
$k d^{2}/n$\\
\hline
% -------------------
$[[26,6,5]]$ & $1 + x^{-2} y^{2} - x^{-2} y^{-1}$ & $x^{2} y^{2} + 1 + x^{-1} y$ & $(0,13)$ & $(1,8)$ & $5.77$ \\
%%%
$[[78,6,11]]$ & $x^{2} y^{-2} + x + 1$ & $-x y^{2} + 1 + x^{-2} y^{-1}$ & $(0,13)$ & $(3,-9)$ & $9.31$ \\
%%%%
$[[104,6,14]]$ & $-x^{2} y^{2} + 1 + x^{-1} y^{-1}$ & $-y^{2} + 1 - y^{-3}$ & $(0,26)$ & $(2,-7)$ & $11.31$ \\
%%%
$[[130,6,16]]$ & $-x^{2} - x^{2} y^{-3} + 1$ & $x^{2} y^{2} + 1 - x^{-1} y^{-2}$ & $(0,13)$ & $(5,-11)$ & $11.82$ \\
%%%
$[[144,6,16]]$ & $xy^{2}+xy+1$ & $xy^{3}+1+x^{-1}y^{3}$ & $(0,24)$ & $(3,-12)$ & $10.67$ \\
%%%%%
$[[156,6,18]]$ & $-xy+1-x^{-2}y^{-3}$ & $1-x^{-1}y+x^{-1}y^{-3}$ & $(0,78)$ & $(1,-59)$ & $12.46$ \\
%%%%
$[[182,6,20]]$ & $y+1+x^{-1}y^{3}$ & $-x^{2}y^{2}-xy^{2}+1$ & $(0,13)$ & $(7,10)$ & $13.19$ \\
%%%
$[[208,6,22]]$ & $xy^{3}-y^{2}+1$ & $-xy^{-2}+1-x^{-1}$ & $(0,13)$ & $(8,6)$ & $13.96$ \\
%%%%
$[[224,6,23]]$ & $x^{3}y^{-1}+1+x^{-1}y^{-3}$ & $1-y^{-1}+x^{-3}y^{-3}$ & $(0,56)$ & $(2,8)$ & $14.17$ \\
%%%
$[[234,6,24]]$ & $-x^{2}y^{-1}+1-x^{-1}y^{-2}$ & $-xy^{-2}+1-x^{-2}y^{-1}$ & $(0,39)$ & $(3,11)$ & $14.77$ \\
% ---------------------------------------------------------------
%\hline\hline
\end{tabular}
\end{table}
%%%%%%%%%%%%%%%%%%%%%%%%%%%%%%%%%
%%%%%%%%%q=5, k=4
%%%%%%%%%%%%%%%%%%%%%%%%%%%%%%%%
\begin{table}[htbp]
\centering
\caption{Same as Table~\ref{tab:3q4k}, but with $q=5$.}
\label{tab:5q4k}
\begin{tabular}{c c c c c c}
\hline\hline
$[[n,k,d]]_{q=5}$ &
\begin{tabular}{@{}c@{}} $f(x,y) =$ \\  \end{tabular} &
\begin{tabular}{@{}c@{}} $g(x,y) =$ \\ \end{tabular} &
$\vec a_1$ &
$\vec a_2$ &
$k d^{2}/n$ \\
\hline
% -------------------
%$[[16,4,4]]$ & $3x^{3}y + 2x^{2} + 1$ & $1 - x^{-1}y^{-2} + 2x^{-3}y^{-2}$ & $(0,2)$ & $(4,1)$ & $4.00$ \\
%%%%
%$[[20,4,5]]$ & $-x y^{2} + 1 + 3x^{-2} y^{2}$ & $3x y^{3} + 1 - x^{-3} y^{2}$ & $(0,2)$ & $(5,1)$ & $5.00$ \\
%%%
$[[24,4,6]]$ & $-x y^{2} + 1 + 3x^{-2}$ & $1 - x^{-2} y^{3} + 2x^{-2} y$ & $(0,4)$ & $(3,1)$ & $6.00$ \\
%%%%%
$[[30,4,7]]$ & $1 + 2x^{-1} y^{2} + 2x^{-1} y^{-1}$ & $1 + y^{-1} + 3x^{-2} y^{-2}$ & $(0,5)$ & $(3,1)$ & $6.53$ \\
%%%%
$[[40,4,9]]$ & $2y + 1 + 2x^{-1} y^{-1}$ & $2x y^{-2} + 1 + 2x^{-2} y^{2}$ & $(0,10)$ & $(2,1)$ & $8.10$ \\
%%%%
$[[48,4,10]]$ & $-x^{2} y^{-1} + 1 + x^{-3} y$ & $2x y + 1 + x^{-1}$ & $(0,8)$ & $(3,4)$ & $8.33$ \\
%%%%
$[[60,4,11]]$ & $3x^{2}y^{-2}+1+x^{-1}y^{-3}$ & $2x^{3}y^{-3}+2x^{2}y+1$ & $(0,6)$ & $(5,-3)$ & $8.07$ \\
%%%%
$[[72,4,13]]$ & $-x^{3}y^{-3}+1+3x^{-2}y^{-2}$ & $3x^{3}y^{-2}+1+2y^{-3}$ & $(0,18)$ & $(2,-3)$ & $9.39$ \\
%%%%
$[[84,4,14]]$ & $x+1+x^{-2}y$ & $1+x^{-1}y^{3}+x^{-1}y^{-1}$ & $(0,6)$ & $(7,-1)$ & $9.33$ \\
%%%%%
$[[90,4,16]]$ & $x^{3} y + 1 + x^{-1} y^{-2}$ & $x^{3} y^{-2} + 1 + x^{-2}$ & $(0,9)$ & $(5,8)$ & $11.38$ \\
%%%%
$[[96,4,16]]$ & $x^{3}y^{-1}+1-y^{-2}$ & $3xy^{-3}+1+2x^{-1}y^{-2}$ & $(0,6)$ & $(8,-2)$ & $10.67$ \\
%%%
$[[108,4,16]]$ & $y+1-x^{-1}y^{-1}$ & $-xy^{-2}+1+x^{-2}y^{2}$ & $(0,27)$ & $(2,-9)$ & $9.48$ \\
%%%
$[[114,4,19]]$ & $1 + 2x^{-3} y + 2x^{-3} y^{-1}$ & $1 + x^{-1} y^{-2} + x^{-2} y^{2}$ & $(0,19)$ & $(3,4)$ & $12.67$ \\
%%%
$[[120,4,19]]$ & $-x^{3}y+1+3x^{-2}y^{3}$ & $x^{3}+2x^{3}y^{-2}+1$ & $(0,15)$ & $(4,-4)$ & $12.03$ \\
%%%%%
$[[132,4,19]]$ & $3x^{3}y^{3}+1+x^{-3}$ & $x^{3}y^{-1}+1+x^{-2}y^{3}$ & $(0,33)$ & $(2,-17)$ & $10.94$ \\
%%%%
$[[144,4,22]]$ & $x^{3}y^{-2}+1+3x^{-2}y^{2}$ & $1+x^{-1}y^{3}+3x^{-1}y^{2}$ & $(0,18)$ & $(4,2)$ & $13.44$ \\
%%%%
$[[168,4,23]]$ & $2x^{2}y^{-1}+1+2y^{-2}$ & $3x^{3}y^{-2}+2x^{2}y^{-2}+1$ & $(0,42)$ & $(2,15)$ & $12.60$ \\
%%%%
$[[204,4,28]]$ & $1+y^{-2}+x^{-1}y^{-3}$ & $1+2y^{-3}+2x^{-3}y^{-3}$ & $(0,51)$ & $(2,13)$ & $15.37$ \\
%%%%
$[[224,4,31]]$ & $3x^{3}y^{-2}+2xy^{-1}+1$ & $2x^{3}y+3x^{2}y^{-3}+1$ & $(0,56)$ & $(2,14)$ & $17.16$ \\
%%%%
$[[246,4,35]]$ & $1+x^{-1}y^{-2}+x^{-2}y^{-1}$ & $2x^{3}y^{-2}+2y+1$ & $(0,41)$ & $(3,-11)$ & $19.92$ \\
% ---------------------------------------------------------------
%\hline\hline
\end{tabular}
\end{table}
%%%%%%%%%%%%%%%%%%%%%%%%%%%%%%%%%%%%%%%%
%%%%%%%%%q=5, k=6
%%%%%%%%%%%%%%%%%%%%%%%%%%%%%%%%%%%%%%%%
\begin{table}[htbp]
\centering
\caption{Same as Table~\ref{tab:3q4k}, but with $q=5$ and $k=6$.}
\label{tab:5q6k}
\begin{tabular}{c c c c c c}
\hline\hline
$[[n,k,d]]_{q=5}$ &
\begin{tabular}{@{}c@{}} $f(x,y) =$ \\  \end{tabular} &
\begin{tabular}{@{}c@{}} $g(x,y) =$ \\ \end{tabular} &
$\vec a_1$ &
$\vec a_2$ &
$k d^{2}/n$ \\
\hline
% -------------------
$[[48,6,8]]$ & $2x^{3}y^{3}+1-x^{-1}y^{3}$ & $-x^{3}y^{-1}+1+3x^{-2}$ & $(0,8)$ & $(3,-6)$ & $8.00$ \\
%%%%
$[[80,6,12]]$ & $3x^{3}y + 1 + x^{-1}$ & $1 + 3x^{-1}y^{3} + x^{-1}y^{2}$ & $(0,20)$ & $(2,7)$ & $10.80$ \\
%%%%
$[[96,6,14]]$ & $-x^{3} y^{-2} + 3x^{2} y^{-2} + 1$ & $3y^{2} + 1 + 2x^{-1} y$ & $(0,6)$ & $(8,5)$ & $12.25$ \\
%%%%
$[[124,6,17]]$ & $3x^{3} y^{2} + x^{2} y + 1$ & $3x^{2} y^{-2} + 3y + 1$ & $(0,31)$ & $(2,21)$ & $13.98$ \\
%%%%
$[[144,6,20]]$ & $1 + y^{-1} + 3x^{-2} y^{3}$ & $3x y^{-2} + 1 + 2x^{-2}$ & $(0,9)$ & $(8,4)$ & $16.67$ \\
%%%
$[[160,6,21]]$ & $xy+1+3x^{-3}y^{3}$ & $3xy^{3}+1+x^{-2}y^{-3}$ & $(0,8)$ & $(10,-4)$ & $16.54$ \\
%%%
$[[168,6,20]]$ & $x^{2}y^{-2}+2x+1$ & $3x^{3}y^{-1}-y+1$ & $(0,28)$ & $(3,9)$ & $14.29$ \\
%%%%
$[[186,6,23]]$ & $2x^{3}y^{-2}+1-x^{-1}y^{-1}$ & $x^{3}y^{2}+1-x^{-2}y^{2}$ & $(0,31)$ & $(3,-18)$ & $17.06$ \\
%%%
$[[192,6,25]]$ & $2xy^{2}+1+x^{-1}y$ & $3x^{2}y^{3}+3xy^{-3}+1$ & $(0,24)$ & $(4,-4)$ & $19.53$ \\
%%%
$[[224,6,26]]$ & $1+3x^{-2}y^{-2}+x^{-3}y^{-1}$ & $-x^{3}y^{3}-x^{3}+1$ & $(0,28)$ & $(4,-8)$ & $18.11$ \\
%%%%
$[[240,6,32]]$ & $3xy^{-1}+1+3x^{-3}y$ & $1-x^{-1}+3x^{-2}y^{-3}$ & $(0,15)$ & $(8,6)$ & $25.60$ \\
% ---------------------------------------------------------------
%\hline\hline
\end{tabular}
\end{table}
%%%%%%%%%%%%%%%%%%%%%%%%%%%%%%%%%%%%%%%%
%%%%%%%%%k=4, q=7
%%%%%%%%%%%%%%%%%%%%%%%%%%%%%%%%%%%%%%%%
\begin{table}[htbp]
\centering
\caption{Same as Table~\ref{tab:3q4k}, but with $q=7$.}
\label{tab:7q4k}
\begin{tabular}{c c c c c c}
\hline\hline
$[[n,k,d]]_{q=7}$ &
\begin{tabular}{@{}c@{}} $f(x,y) =$ \\  \end{tabular} &
\begin{tabular}{@{}c@{}} $g(x,y) =$ \\ \end{tabular} &
$\vec a_1$ &
$\vec a_2$ &
$k d^{2}/n$ \\
\hline
% -------------------
%%%%
$[[24,4,6]]$ & $5x^{3} y^{2} - x^{2} y + 1$ & $2x + 1 + 2y^{-1}$ & $(0,6)$ & $(2,3)$ & $6.00$ \\
%%%
$[[28,4,7]]$ & $3x^{2} y^{-1} + 4x y^{-1} + 1$ & $1 - x^{-1} y^{-2} + 2x^{-3} y^{3}$ & $(0,7)$ & $(2,-6)$ & $7.00$ \\
%%%%
$[[32,4,7]]$ & $2x y^{-3} + 1 + 5x^{-1} y^{-1}$ & $4x + 2x y^{-1} + 1$ & $(0,4)$ & $(4,-3)$ & $6.13$ \\
%%%
$[[48,4,10]]$ & $1 - x^{-2} y^{-2} + 4x^{-3}$ & $1 - x^{-2} + 3x^{-3} y^{-3}$ & $(0,6)$ & $(4,-1)$ & $8.33$ \\
%%%%
$[[56,4,11]]$ & $x^{3} y + 5y^{2} + 1$ & $5x^{2} y^{3} + 3y + 1$ & $(0,28)$ & $(1,-15)$ & $8.64$ \\
%%%%
$[[64,4,12]]$ & $5xy^{2} + 2xy + 1$ & $5xy^{3} + 1 + 2x^{-1}y^{3}$ & $(0,8)$ & $(4,-6)$ & $9.00$ \\
%%%%
$[[96,4,17]]$ & $3x^{3}y^{-3}+2y^{2}+1$ & $3x^{2}y+1+3x^{-1}y^{-3}$ & $(0,48)$ & $(1,8)$ & $12.04$ \\
%%%%
$[[120,4,20]]$ & $4xy+4xy^{-1}+1$ & $5x^{3}y+1+4y^{-3}$ & $(0,30)$ & $(2,-8)$ & $13.33$ \\
%%%%
$[[168,4,27]]$ & $x^{3}y^{2}+1+y^{-3}$ & $2x^{2}y^{3}+2x^{2}y+1$ & $(0,21)$ & $(4,-7)$ & $17.36$ \\
%%%%
$[[192,4,32]]$ & $2x^{2}y^{-1}+1+4y^{-3}$ & $2x^{2}y+3xy^{3}+1$ & $(0,96)$ & $(1,-9)$ & $21.33$ \\
%%%
% ---------------------------------------------------------------
%\hline\hline
\end{tabular}
\end{table}
%%%%%%%%%%%%%%%%%%%%%%%%%%%%%%%%%%%%%%%%
%%%%%%%%%table
%%%%%%%%%%%%%%%%%%%%%%%%%%%%%%%%%%%%%%%%
\begin{table}[htbp]
\centering
\caption{Same as Table~\ref{tab:3q4k}, but with $q=7$ and $k=6$.}
\label{tab:7q6k}
\begin{tabular}{c c c c c c}
\hline\hline
$[[n,k,d]]_{q=7}$ &
\begin{tabular}{@{}c@{}} $f(x,y) =$ \\  \end{tabular} &
\begin{tabular}{@{}c@{}} $g(x,y) =$ \\ \end{tabular} &
$\vec a_1$ &
$\vec a_2$ &
$k d^{2}/n$ \\
\hline
% -------------------
%$[[48,6,7]]$ & $4xy + 1 - x^{-2} y^{-2}$ & $4xy^{2} + 1 + x^{-2} y^{2}$ & $(0,12)$ & $(2,-9)$ & $6.13$ \\
%%%
$[[76,6,11]]$ & $2x^{2} y^{3} + 1 - x^{-3} y^{3}$ & $1 + y^{-2} + x^{-3} y^{-1}$ & $(0,19)$ & $(2,10)$ & $9.55$ \\
%%%%
$[[96,6,15]]$ & $3x y^{2} + y + 1$ & $2x^{3} y^{-3} + 3x + 1$ & $(0,24)$ & $(2,7)$ & $14.06$ \\
%%%%%
$[[108,6,15]]$ & $-x^{2}y+1+2x^{-2}y^{-2}$ & $1+4y^{-2}+2x^{-3}y^{-3}$ & $(0,18)$ & $(3,-12)$ & $12.50$ \\
%%%%
$[[126,6,18]]$ & $xy^{-3}+1+5y^{-3}$ & $1+y^{-2}+x^{-2}y^{2}$ & $(0,21)$ & $(3,-12)$ & $15.43$ \\
%%%
$[[128,6,18]]$ & $-x + 1 + 2x^{-3} y^{-2}$ & $2x y^{3} + 3y + 1$ & $(0,16)$ & $(4,-4)$ & $15.19$ \\
%%%%
$[[152,6,20]]$ & $1+x^{-2}y^{-1}-x^{-3}y^{-1}$ & $2xy^{-2}+1+x^{-3}y$ & $(0,19)$ & $(4,-11)$ & $15.79$ \\
%%%%
$[[180,6,25]]$ & $3x^{3}y^{3}+1+5x^{-2}y^{-3}$ & $2x^{3}y^{-1}+1+x^{-1}y$ & $(0,30)$ & $(3,12)$ & $20.83$ \\
%%%
$[[228,6,29]]$ & $1+4x^{-1}y^{-2}+2x^{-2}y^{-1}$ & $x^{3}y^{-2}+4y+1$ & $(0,114)$ & $(1,-11)$ & $22.13$ \\
% ---------------------------------------------------------------
%\hline\hline
\end{tabular}
\end{table}
%%%%%%%%%%%%%%%%%%%%%%%%%%%%%%%%%
%%%%%%%%%
%%%%%%%%%%%%%%%%%%%%%%%%%%%%%%%%

\section{Conclusions and Outlook}
\label{sec:conclusion}

In this work we investigated finite-length qudit quantum LDPC codes obtained
from translation-invariant CSS constructions on twisted tori, motivated by the
recent observation that twisted boundary identifications can strongly improve
finite-size behavior. Using an algebraic description over finite fields, we
computed the number of logical qudits efficiently for large collections of
candidate patterns and combined this with randomized distance estimation to
identify good finite codes within a sparse, weight-$6$ ansatz. The best
instances found are summarized in Tables~\ref{tab:3q4k}--\ref{tab:7q6k}.
On the theory side we made explicit what changes when passing from qubits
to qudits: the relative sign required in the check generators for odd $q$,
the coefficient freedom in $\mathbb{F}_q^{\times}$, and the anyon periods
governed by $q^{m}-1$ for zeros in the extension field $\mathbb{F}_{q^m}$,
which explain why twisted tori are particularly useful for qudits.

Our numerical results indicate that twisted-torus qudit generalized toric
constructions can substantially improve the finite-length figure of merit
$k d^2/n$ compared to untwisted geometries, and in the regime explored they also
compare favorably with previously reported twisted-torus qubit instances.
Together, these observations suggest that the combination of higher local
dimension and twisted boundary conditions provides a practical lever for
improving finite-size rate--distance tradeoffs in translation-invariant LDPC
families. One limitation should be kept in mind: the reported distances are
upper bounds from randomized estimation.

Several directions remain open. First, it would be interesting to extend the
search beyond the present sparse ansatz, for example by allowing larger support
or additional structure (e.g.\ self-duality constraints) while keeping checks
bounded-weight. Second, improving distance certification---either by stronger
randomized routines or by exact methods for selected parameter ranges---would
sharpen comparisons across code families. Third, it would be valuable to study
decoding performance under realistic noise models and to identify twists that
optimize not only $(n,k,d)$ but also decoding thresholds and runtime. Finally,
our results motivate a more systematic understanding of how twisted
identifications interact with the algebraic data $(f,g)$ to control logical
dimension and distance in the qudit setting.

\paragraph*{Code availability.}
The SageMath and GAP code used for the numerical search and the distance
estimation is available at
\url{https://github.com/halmr/qudit-twisted-torus-codes}.

%%%%%%%%%%%%%%%%%%%%%%%%%%%%%%%%%%%%%%%%%%%%%%%%%%%%%%%%%%%%%%%%%%%%%%%%%%%%%%%%%%
%appendices
%%%%%%%%%%%%%%%%%%%%%%%%%%%%%%%%%%%%%%%%%%%%%%%%%%%%%%%%%%%%%%%%%%%%%%%%%%%%%%%%%
\appendix

\section{Numerical procedure used to produce the tables}
\label{app:algorithm}

This appendix describes how the entries of
Tables~\ref{tab:3q4k}--\ref{tab:7q6k} were obtained, so that the search can
be repeated.  There are three steps: the logical dimension $k$ of a
candidate code on a given torus is computed with a Gr\"obner basis
(Algorithm~\ref{alg:k}); a search runs over candidate polynomials $(f,g)$
and tori $(\alpha,\beta,\gamma)$, keeps the candidates with the target $k$,
and estimates their distance (Algorithm~\ref{alg:search}); for each
$(n,k,q)$ the best candidate found in the search is placed in the table.
For $p\in R_q$ we write $\tilde p$ for $p$ multiplied by the smallest
monomial that removes its negative exponents; this does not change the
ideal generated in $R_q$.

\paragraph{Computing $k$ on a torus.}
Algorithm~\ref{alg:k} evaluates \eqref{eq:k_twisted_qudit_short}.  The
ideal is formed in the polynomial ring $\mathbb{F}_q[x,y]$, with negative
exponents cleared separately in each generator.  The saturation step
removes solutions with $x=0$ or $y=0$, which exist in $\mathbb{F}_q[x,y]$
but not in the Laurent ring.  The number of standard monomials of the
resulting ideal equals the dimension of the quotient
$\mathcal{M}(\alpha,\beta,\gamma)$, and $k$ is twice this number.  The
value $k_{\max}$ of \eqref{eq:kmax_qudit} is obtained by the same steps
without the two boundary relations.

\begin{algorithm}[H]
\caption{Logical dimension on a twisted torus}
\label{alg:k}
\begin{algorithmic}[1]
\Require prime $q$; $f,g\in R_q$; torus $(\alpha,\beta,\gamma)$
\Ensure $k(\alpha,\beta,\gamma)$
\State $J\gets\langle\,\tilde f,\;\tilde g,\;y^{\alpha}-1,\;x^{\beta}y^{\gamma\bmod\alpha}-1\,\rangle\subset\mathbb{F}_q[x,y]$
\State $J\gets\bigl(J:(xy)^{\infty}\bigr)$ \Comment{remove solutions with $x=0$ or $y=0$}
\State $G\gets$ Gr\"obner basis of $J$
\If{$G=\{1\}$} \State \Return $0$ \EndIf
\State $B\gets$ normal basis of $J$ (monomials not divisible by any leading monomial of $G$)
\State \Return $2\,|B|$
\end{algorithmic}
\end{algorithm}

\paragraph{The search.}
Algorithm~\ref{alg:search} generates candidate pairs $(f,g)$ of the form
\eqref{ansatz} and keeps those that are coprime (the topological-order
condition \eqref{eq:TO_ideal_qudit}), have $k_{\max}\ge k$, and have
exactly the target $k$ on a given torus.  For these, the check matrices
$H_X$ and $H_Z$ are built from the shift matrices $X,Y$ of the torus; we
require $H_XH_Z^{\mathsf T}=0$ and that every row of $H_X$ and $H_Z$ has
weight $t_f+t_g$, which excludes tori on which two monomials of $f$ or of
$g$ fall on the same qudit.  The distance is estimated with
\textsc{DistRandCSS} of the GAP package QDistRnd \cite{QDistRnd}, applied
to $(H_X,H_Z)$ and to $(H_Z,H_X)$; it returns the smallest weight of a
logical operator found with $N_{\mathrm{is}}$ random information sets, so
$d$ is an upper bound on the true distance.

\begin{algorithm}[H]
\caption{Search for twisted-torus codes with target parameters $(n,k)$}
\label{alg:search}
\begin{algorithmic}[1]
\Require prime $q$; target $(n,k)$; ansatz \eqref{ansatz} with weights
         $t_f,t_g$ and exponent set $\mathcal{E}\subset\mathbb{Z}^2$;
         twist set $\mathcal{T}\subset\mathbb{Z}$; number of information sets $N_{\mathrm{is}}$
\Ensure codes $[[n,k,d]]_q$ with $d$ an upper bound on the distance
\State $\mathcal{L}\gets\emptyset$
\For{each pair $(f,g)$ of the form \eqref{ansatz} with distinct exponents in $\mathcal{E}$ and coefficients in $\mathbb{F}_q^{\times}$, enumerated or sampled}
  \If{$\gcd(\tilde f,\tilde g)$ is not a monomial} \State \textbf{continue} \Comment{condition \eqref{eq:TO_ideal_qudit}} \EndIf
  \If{$k_{\max}<k$} \State \textbf{continue} \EndIf
  \For{each $(\alpha,\beta)$ with $2\alpha\beta=n$}
    \If{$k(\alpha,\beta,0)>k$, or the untwisted torus $\gamma=0$ fails the checks $H_XH_Z^{\mathsf T}=0$ and row weight $t_f+t_g$} \State \textbf{continue} \Comment{untwisted torus evaluated first} \EndIf
    \For{each $\gamma\in\mathcal{T}$ with $|\gamma|<\alpha$}
      \If{$k(\alpha,\beta,\gamma)\neq k$} \State \textbf{continue} \Comment{Algorithm~\ref{alg:k}} \EndIf
      \State $A\gets f(X,Y)$, $B\gets g(X,Y)$, with $X,Y$ the shift matrices of the torus \eqref{eq:torus_relations}
      \State $H_X\gets[\,A\;|\;B\,]$,\quad $H_Z\gets[\,B^{\mathsf T}\;|\;-A^{\mathsf T}\,]$
      \If{$H_XH_Z^{\mathsf T}\neq0$, or some row of $H_X$ or $H_Z$ has weight $\neq t_f+t_g$} \State \textbf{continue} \EndIf
      \State $d\gets\min\{\textsc{DistRandCSS}(H_X,H_Z,N_{\mathrm{is}}),\ \textsc{DistRandCSS}(H_Z,H_X,N_{\mathrm{is}})\}$
      \State add $(f,g,\alpha,\beta,\gamma,k,d)$ to $\mathcal{L}$
    \EndFor
  \EndFor
\EndFor
\State \Return the element of $\mathcal{L}$ with the largest $kd^{2}/n$; ties broken by the smallest stabilizer range
\end{algorithmic}
\end{algorithm}

\paragraph{Selection and parameters.}
For each $(n,k,q)$ the table lists the candidate with the largest
$kd^{2}/n$ found; if several candidates reach the same value, the one with
the smallest stabilizer range (Definition~\ref{def:range}) is listed.  The
tables were produced with $t_f=t_g=3$, exponents in $\{-3,\dots,3\}$, all
factorizations $n=2\alpha\beta$ with $\alpha,\beta\ge1$, all twists
$\gamma$ with $|\gamma|<\alpha$, the number of information sets stated in
Sec.~\ref{sec:numerics}, and between $1000$ and $10\,000$ randomly sampled
pairs $(f,g)$ per target $(n,k)$, depending on $n$, with a fixed random
seed.  For each $(\alpha,\beta)$ the untwisted torus $\gamma=0$ is
evaluated first, and the pair is skipped if its untwisted logical dimension
exceeds the target $k$ or if it fails the weight and commutation checks.

\paragraph{Remark.}
With $q=2$ the same procedure reproduces the parameters of the codes of
Ref.~\cite{LiangTwistedTori} that we tested.


\newpage
\begin{thebibliography}{99}

%---------------------------------------------------------------------
% [D] BIBLIOGRAPHY  -- two new entries (please verify)
%---------------------------------------------------------------------
%1
\bibitem{KitaevToric}
A. Yu. Kitaev, ``Fault-tolerant quantum computation by anyons,''
Annals of Physics, 303, 2-30 (2003).
\url{https://doi.org/10.1016/S0003-4916(02)00018-0}.
%2
\bibitem{BombinHomological}
H. Bombin and M. A. Martin-Delgado, ``Homological error correction: Classical and quantum codes,''
Journal of Mathematical Physics, 48, 052105 (2007).
\url{https://doi.org/10.1063/1.2731356}.
%3
\bibitem{BreuckmannEberhardtLDPC}
N. P. Breuckmann and J. N. Eberhardt, ``Quantum Low-Density Parity-Check Codes,''
PRX Quantum, 2, 040101 (2021).
\url{https://doi.org/10.1103/PRXQuantum.2.040101}.
%4
\bibitem{TillichZemor}
J.-P. Tillich and G. Z\'emor, ``Quantum LDPC codes with positive rate and minimum distance proportional to the square root of the blocklength,''
IEEE Transactions on Information Theory, 60, 1193-1202 (2014).
\url{https://doi.org/10.1109/TIT.2013.2292061}.
%5
\bibitem{BravyiBB}
S. Bravyi, A. W. Cross, J. M. Gambetta, D. Maslov, P. Rall, and T. J. Yoder,
``High-threshold and low-overhead fault-tolerant quantum memory,''
Nature, 627, 778-782 (2024).
\url{https://doi.org/10.1038/s41586-024-07107-7}.
%6
\bibitem{HaahModules}
J. Haah, ``Commuting Pauli Hamiltonians as maps between free modules,''
Communications in Mathematical Physics, 324, 351-399 (2013).
\url{https://doi.org/10.1007/s00220-013-1810-2}.
%7
\bibitem{LiangTopOrder}
Z. Liang, Y. Xu, J. T. Iosue, and Y.-A. Chen,
``Extracting Topological Orders of Generalized Pauli Stabilizer Codes in Two Dimensions,''
PRX Quantum, 5, 030328 (2024).
\url{https://doi.org/10.1103/PRXQuantum.5.030328}.
%8
\bibitem{LiangTwistedTori}
Z. Liang, K. Liu, H. Song, and Y.-A. Chen,
``Generalized Toric Codes on Twisted Tori for Quantum Error Correction,''
PRX Quantum, 6, 020357 (2025).
\url{https://doi.org/10.1103/rmy6-9n89}.
%9
\bibitem{LiangPlanar}
Z. Liang, J. N. Eberhardt, and Y.-A. Chen,
``Planar Quantum Low-Density Parity-Check Codes with Open Boundaries,''
PRX Quantum, 6, 040330 (2025).
\url{https://doi.org/10.1103/qv65-vmzr}.
%10
\bibitem{BullockBrennenQuditSurface}
S. S. Bullock and G. K. Brennen, ``Qudit surface codes and gauge theory with finite cyclic groups,''
Journal of Physics A: Mathematical and Theoretical, 40, 3481 (2007).
\url{https://doi.org/10.1088/1751-8113/40/13/013}.
%11
\bibitem{SpencerQuditLDPC}
D. J. Spencer, A. Tanggara, T. Haug, D. Khu, and K. Bharti,
``Qudit low-density parity-check codes,''
Quantum, 10, 2023 (2026).
\url{https://doi.org/10.22331/q-2026-03-13-2023}.
%12
\bibitem{SageMath}
The Sage Developers, \textit{SageMath, the Sage Mathematics Software System}, version~10.9,
\url{https://www.sagemath.org}.
%13
\bibitem{QDistRnd}
L. P. Pryadko, V. A. Shabashov, and V. K. Kozin,
``QDistRnd: A GAP package for computing the distance of quantum error-correcting codes,''
Journal of Open Source Software, 7(71), 4120 (2022).
\url{https://doi.org/10.21105/joss.04120}.
%14
\bibitem{HaahQudit}
J. Haah, ``Classification of translation invariant topological Pauli
stabilizer codes for prime dimensional qudits on two-dimensional lattices,''
Journal of Mathematical Physics, 62, 012201 (2021).
%15
\bibitem{KetkarNonbinary}
A. Ketkar, A. Klappenecker, S. Kumar, and P. K. Sarvepalli,
``Nonbinary stabilizer codes over finite fields,''
IEEE Transactions on Information Theory, 52, 4892--4914 (2006).
\url{https://doi.org/10.1109/TIT.2006.883612}.
%16
\bibitem{LidlNiederreiter}
R. Lidl and H. Niederreiter, \emph{Introduction to Finite Fields and Their
Applications} (Cambridge University Press, Cambridge, 1986).
%17
\bibitem{TerhalBound1}
S. Bravyi, B. Terhal, ``A no-go theorem for a two-dimensional self-correcting quantum memory based on stabilizer codes,''
New J. Phys., 11, 043029 (2009).
\url{https://doi.org/10.1088/1367-2630/11/4/043029}.
%18
\bibitem{TerhalBound2}
S. Bravyi, D. Poulin, and B. Terhal, ``Tradeoffs for reliable quantum information storage in 2D systems,''
Phys. Rev. Lett., 104, 050503 (2010).
\url{https://doi.org/10.1103/PhysRevLett.104.050503}.




\end{thebibliography}
\end{document}